\documentclass[aps,twocolumn,showpacs,fleqn]{revtex4}

\usepackage{graphicx}
\usepackage{amsmath}
\usepackage{amssymb}

\newcommand{\eq}[1]{Eq.\,(\ref{#1})} 
 
\newcommand{\fig}[1]{Fig.\,\ref{#1}} 
 
\renewcommand{\vec}[1]{\mathbf #1}

\def\eps{\varepsilon}
\def\d#1{\mathrm{d}#1}

\begin{document}
\title{Many-electron dynamics triggered by massively parallel ionization}
\author{Christian Gnodtke$^1$, Ulf Saalmann$^{1,2}$, and Jan-Michael Rost$^{1,2}$} 
\affiliation{$^1$Max Planck Institute for the Physics of Complex Systems, 
N\"othnitzer Stra{\ss}e 38, 01187 Dresden, Germany\\
$^2$Max Planck Advanced Study Group at CFEL, Luruper Chaussee 149, 22761 Hamburg, Germany} 
\date{\today}

\begin{abstract}\noindent 
Massively parallel ionization of many atoms in a cluster or bio-molecule is identified as new phenomenon
of light-matter interaction which becomes feasible through short and intense FEL pulses.
Almost simultaneously emitted from the illuminated target the photo-electrons can have such a high density that they interact substantially even {\em after} photoionization. 
This interaction results in a characteristic electron spectrum which can be interpreted as convolution of a mean-field electron dynamics and binary electron-electron collisions. We demonstrate that this universal spectrum can be obtained analytically by summing synthetic two-body Coulomb collision events. Moreover, we propose an experiment with hydrogen clusters to observe massively parallel ionization.
\end{abstract}


\maketitle

\noindent 
Modern light sources such as free-electrons lasers (FELs) \cite{acas+07,emak+10} couple large numbers of photons into clusters \cite{sasi+06} or bio-molecules \cite{newo+00}, or more generally, extended systems.
Within femtoseconds many electrons are released through single-photon absorption and the ions left behind form a deep background potential. In cases, where most electrons are trapped in this potential, one observes a sharp transition from continuous equilibration of the photo-activated electrons \cite{both+10,arfe10,gnsa+11,arfe11} to a non-equilibrium plasma executing characteristic oscillations \cite{sage+08} as the pulse length falls below the relaxation time.
If the electrons are activated with sufficient energy to escape the potential in large numbers, a similar transition occurs when the pulse length falls below a critical escape time enabling direct interaction and energy exchange among the electrons even after photo-ionization, as we will show in the following. The regime introduced and discussed here is the exact opposite to the previously investigated case of ``non-interacting electrons'' in sequential emission (also referred to as multi-step ionization \cite{both+08}), which occurs for long pulses \cite{both+08,mo09,gnsa+11}.

The high energy of the excited electronic system permits a treatment in terms of classical Coulomb dynamics of ions and electrons \cite{newo+00,saro02,jufa+04,sasi+06}. This is a tremendous simplification and allows us 
to calculate the time evolution of this many-body system and the photo-electron spectrum (PES) which results from illumination of the cluster with an intense high-frequency laser pulse using classical molecular dynamics with photo-ionization rates for the atomic ionization within the atomic cluster \cite{saro02}. 

We will interpret the PES in two different ways: In terms of global types of dynamics we will show that massively parallel ionization can be thought of as a convolution of a mean-field component and a component typical for binary collisions. In terms of detailed paths of electrons we will demonstrate that the electron spectrum can be reproduced extremely well by approximating each photo-electron's final energy through a sum of contributions from synthetic binary collisions with each of the other electrons. 

%%%%%%%%%%%%%%
\begin{figure}[b]
\begin{center}
\includegraphics[width=\columnwidth]{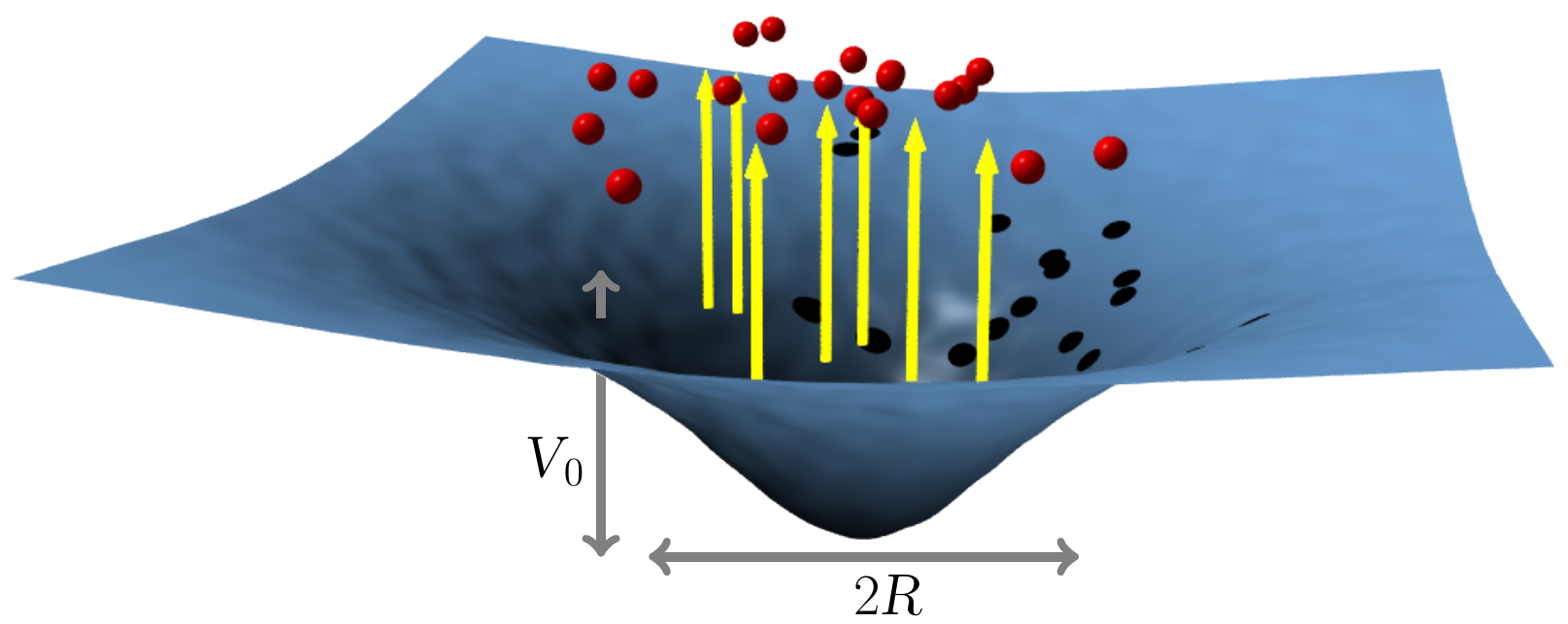}
\end{center}
\caption{Sketch of $N$-fold photo-ionization of atoms in a cluster of radius $R$, leading to an ionic back ground potential (blue) with total charge $N$ and depth $V_{0}=3N/2R$, cf.\ Eq.\,\eqref{eq:pot}.
}
\label{fig:sketch}
\end{figure}%
%%%%%%%%%%%%%
This two-fold interpretation is facilitated by the concept of Coulomb complexes (CC), which we have introduced recently \cite{gnsa+11}. They capture the essentials of electron dynamics activated through multiple photo-ionization in a cluster or bio-molecule. To understand massively parallel ionization, we only need CCs in their simplest version: A single isolated photoionization event from the mother atom in the cluster leads to an excess energy of $E^{*}$. If $N$ photo-electrons are produced by a laser pulse of length $T$ the ions left behind are assumed to form a smooth background potential, see \fig{fig:sketch}. It is Coulombic outside the (spherical) cluster of radius $R$, $V(r) = -N/r$ for $r>R$, and parabolic within the cluster,
\begin{equation}\label{eq:pot}
V(r) = V_{0}\left[r^{2}/(3R^{2})-1\right],\quad r<R
\end{equation}
with the depth $V_{0}\equiv\frac{3}{2}N/R$.
Activated electrons are propagated under this potential and their mutual Coulomb repulsion.
The overall electronic dynamics of the CC is completely determined by the four parameters 
$(N,R,E^{*},T)$. Moreover, CCs are scale invariant, i.e., the one-parameter manifold of 
CC$(\eta) = \{(N,\,\eta R,\,\eta^{-1} E^{*},\,\eta^{-3/2}T)\,|\,\eta{>}0\}$
leads to the same scaled dynamics. On the one hand, this renders phenomena which can be described by CCs quite general and, on the other hand, facilitates to identify a parameter combination which can be realized in an experiment.

\begin{figure}[t]
\begin{center}
\includegraphics[width=\columnwidth]{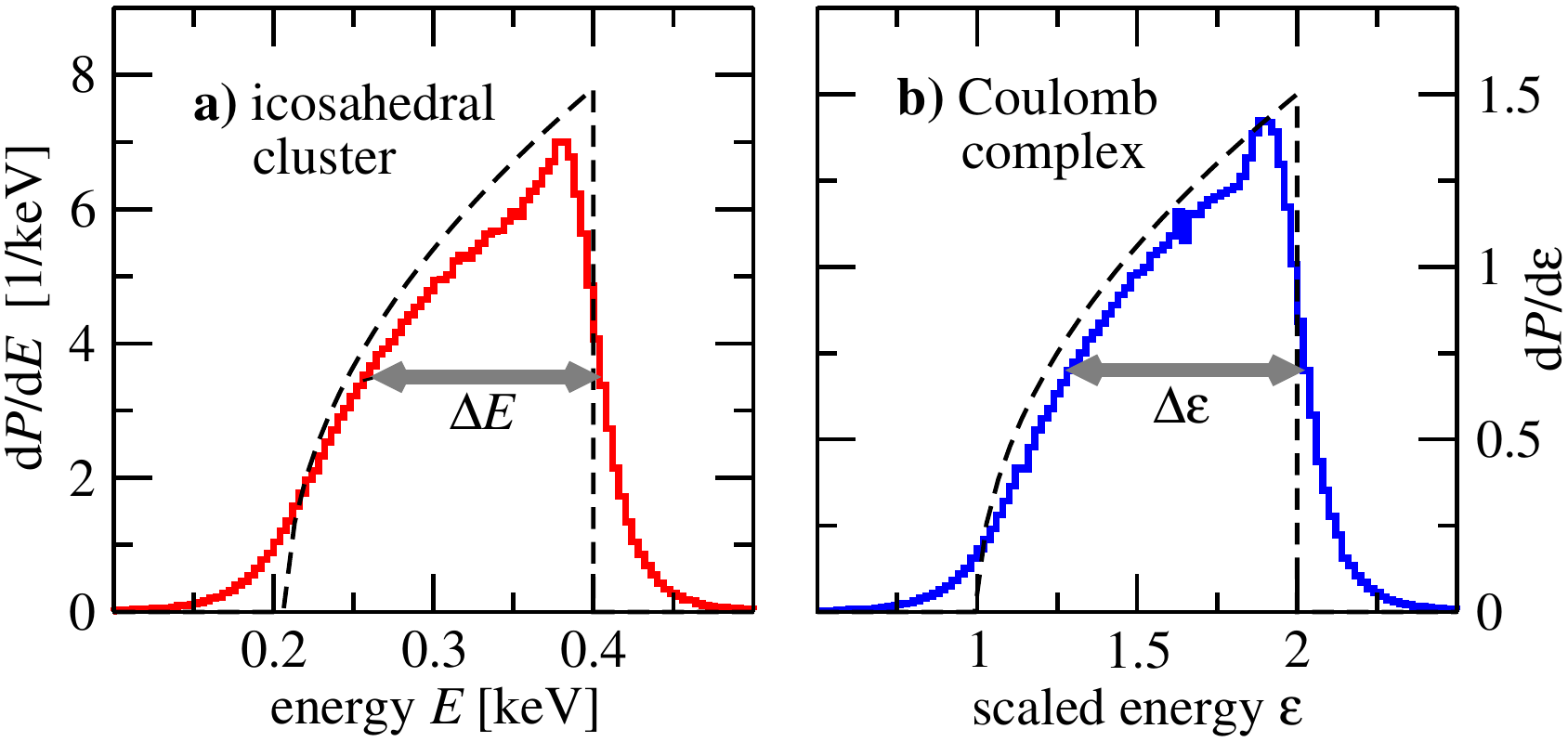}
\end{center}
\caption{Color online: Photo-electron spectrum for sudden massively parallel ionization. a) Ar$_{147}$ with icosahedral geometry. Two thirds of the atoms loose an electron from the 3p level with an excess energy of $E^{\star}\,{=}\,0.4$\,keV. b) Coulomb complex with $N\,{=}\,100$ and an excess energy of $\eps^{*}\,{=}\,2$. The dashed lines represent the analytical mean-field result from Eq.\,\eqref{eq:pes}. The full widths at half maximum of both distributions, denoted by $\Delta E$ and $\Delta\eps$ respectively, are indicated by the thick gray arrows.}
\label{fig:cluscompl}
\end{figure}

In \fig{fig:cluscompl} we compare the PES from the full molecular dynamics calculation (where electrons and ions move classically according to all Coulomb forces) with the one obtained using the CC with its static and smooth ionic background. Obviously, both results agree with each other quite well demonstrating that CCs are a realistic approximation for the present scenario. 
Motivated by the scaling property of the Coulomb complex, we rescale the energy by the depth of the ionic background potential $V_{0} = \frac{3}{2}N/R \equiv \eta$ of all activated electrons potential. In \fig{fig:cluscompl}b the PES is plotted in terms of the scaled energy $\eps = \eta^{-1}E \equiv E/V_{0}$.

The form of the PES exhibits some resemblance to the mean-field result (dashed lines in \fig{fig:cluscompl}) which can be obtained analytically from the CC: Assuming as before spherical geometry, 
the potential for an electron photo activated with excess energy $E^{*}$ at radial distance $r'$ from the center is $V_{\rm mf}(r')= V(r') + V_{\rm ee}(r')$,
where $V$ is given in \eq{eq:pot} and $V_{\rm ee}(r')= N_{r'}/r'$ is the repulsive potential of the charge $N_{r'}=N\,r'^{3}\!/R^{3}$ created by all electrons within the sphere of radius $r'$. To escape from the CC the electron has to overcome
$V_{\rm mf}$ and its final energy is therefore $E=E^{\star}+V_{\rm mf}(r')$ or in scaled units
\begin{equation}\label{eq:finale}
\eps(r') = \eps^* - \left(r'^2/R^2-1\right).
\end{equation} 
% which can be inverted $r'(\eps) = R\sqrt{1+\eps-\eps^*}$. 
With the radial electron distribution $\d{P}/\d{r'} = 3r'^2/R^3$ and \eq{eq:finale} we get for $\d{P}/\d{\eps} = [\d{P}/\d{r'}]\,[\d{\eps} /\d{r'}]^{-1}$
\begin{equation}\label{eq:pes}
\frac{\d{P}(\eps) }{\d{\eps} } =
% R \left( \eps - \eps^* +1 \right)^{1/2},
 \frac{3}{2} \sqrt{\eps - \eps^* +1 },
\end{equation}
within the interval $\eps^{*}-1\le \eps\le \eps^{*}$ which is of length unity or $V_{0}$ in unscaled energies.
The width $V_{0}$ of the PES gives an account of the depth of the potential and consequently of the charge to extension ratio of the Coulomb complex. 
In particular the full width at half maximum
\begin{equation}\label{eq:delte}
 \Delta E=\frac{3}{4}V_{0}=\frac{9}{8}N/R
 \quad\mbox{or}\quad
 \Delta\eps=\frac{3}{4}
\end{equation}
of the mean-field spectrum is quite similar
to its counterpart in the full spectrum.
In the regime of massively parallel ionization this result is very useful to estimate the number of photons absorbed if the cluster size is known, or vice versa, determine the size of the cluster illuminated if one can measure how many electrons (their number equals the number of photons $N$) have been released.

Apart from the overall agreement one observes in \fig{fig:cluscompl} that the accurate PES is substantially blurred at the boundaries compared to the mean-field spectrum. The broadening
 can be interpreted as the result of a convolution with a spectrum governed by binary collisions induced by a \emph{short}-range, singular potential, i.e., the exact opposite of mean-field dynamics, which is generated by smooth \emph{long}-range interaction,
\begin{equation}\label{eq:yukf}
\frac{\d{P}(\eps)}{\d{\eps}} = \int \!\d{\bar{\eps}}\, 
\frac{\d{P}_{\mathrm{long}}(\eps)}{\d{\eps}}\Big|_{\bar{\eps}} 
\frac{\d{P}_{\mathrm{short}}(\eps)}{\d{\eps}}\Big|_{\bar{\eps} - \eps - \eps^*}.
\end{equation}
For the sake of being specific we model the short-range interaction of two electrons with a distance $r$ by a Yukawa potential $W_{\rm short}(r)=\mathrm{e}^{-r/s}/r$
and the mean-field interaction by a Coulomb potential whose singularity at the origin is suppressed, $W_{\rm long}(r)=(1-\mathrm{e}^{-r/s})/r$. % for which $W_{\rm long}(r{\to}0){=}1/s$.
For the screening parameter we choose $s=\frac 15RN^{-1/3}$, i.\,e., much smaller than the initial nearest neighbour distance.
With this choice $W_{\rm long}$ closely matches the Coulombic case, while the initial interaction energy for $W_{\rm short}$ is close to zero. Consequently, the inter-electronic repulsion energy resulting from $W_{\rm short}$ cannot compensate the ionic background potential \eq{eq:pot} anymore. For realistically modeling with $W_{\rm short}$ the effect of binary interactions on the PES, we drop the background potential but adjust the initial conditions such that the asymptotic single electron energy of $\eps^{*}$ is preserved.

%%%%%%%%%%%%%%
\begin{figure}[t]
\begin{center}
\includegraphics[width=0.85\columnwidth]{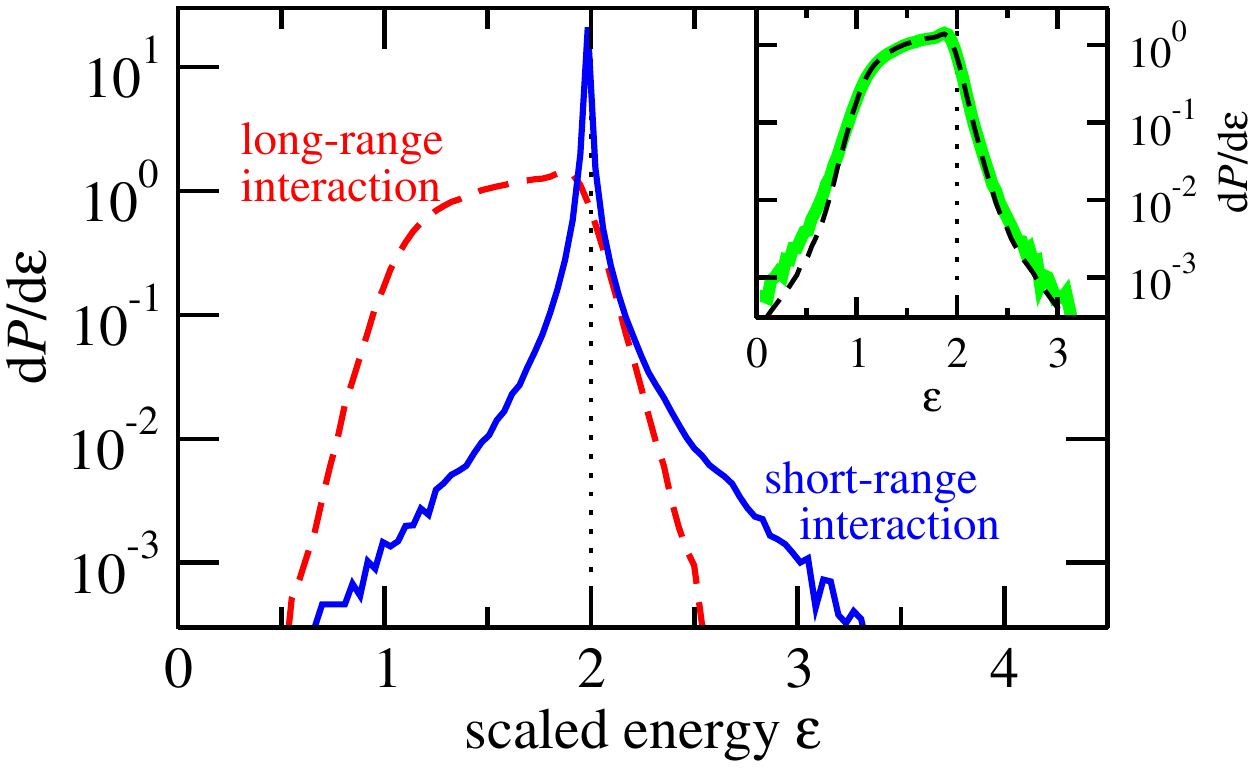}
\end{center}
\caption{Color online: (a) Final electron spectra for $N=10^{2}$ electrons propagated with Yukawa (blue, solid) and anti-Yukawa (red, dashed) potential. (b) Final spectrum from propagation with Coulomb potential (green, solid) and convolution of Yukawa and anti-Yukawa spectra (black, dashed) according to \eq{eq:yukf}. Dotted vertical lines indicate the excess energy $\eps^* = 2$.}
\label{fig:yukawa}
\end{figure}

%%%%%%%%%%%%%
The long-range case gives a final spectrum closely resembling the mean-field PES, albeit already slightly broadened at the edges. The short-range case, on the other hand, leads to a nearly symmetric spectrum sharply peaked at the single electron energy $\eps^*$ but with long tails reaching well beyond energies $\eps=\eps^* \pm 1$. 
Due to the rapid fall-off of the short-range potential most electrons do not interact with each other after photo-absorption. Only if the initial velocity vectors of two electrons put them onto a colliding trajectory an exchange of energy among these two electrons is achieved. Due to the high initial kinetic energy of the order $\eps^*$ only a small subset of initial conditions leads to electron pair trajectories with large energy exchange. Thereby, in a single binary collision one electron can transfer all its kinetic energy to its collision partner and consequently the spectrum in \fig{fig:cluscompl} covers the range $\eps = 0\ldots4$ for $\eps^{\star}=2$. Since such violent binary collisions are very rare on the one hand, but lead to the largest energy exchange on the other hand, they can be viewed as an additional and largely {\it independent} random event, which augments the mean-field expansion dynamics. 
 If this description is realistic, the convolution \eq{eq:yukf} of the short-range and long-range spectrum should reproduce the full spectrum which is indeed the case as shown with the inset of \fig{fig:yukawa}. 
A residual interaction effect manifests itself mostly in slight deviations at low energies: 
While the convolution of \eq{eq:yukf} ascribes each electron the same probability that its final energy $\eps$ gets modified by a violent binary collisions, this is in reality more likely for slow electrons
They come from the central, bulk-like region of the cluster and are more likely to suffer multiple collisions during their escape as compared to surface electron which are faster and suffer at most a single collision.

%%%%%%%%%%%%%%%%%
%%%%%%%%%%%%%%
The described construction of the PES from a mean-field and a binary-interaction component provides an intuitive physical picture. Yet, despite its approximate character it offers no computational advantage over the full result, since all trajectories for $W_{\mathrm{short}}$ must be obtained numerically.
Surprisingly, it is possible to take into account the correlation of mean-field and collision dynamics accurately by determining for each electron its pairwise isolated, binary Coulomb dynamics with all other electrons. This leads to the \emph{binary-intercation sum} (BIS), a quasi-analytical and very accurate formulation for the
PES which we introduce now.

%%%%%%%%%%%%%%%%%
\begin{figure}[b]
\begin{center}
\includegraphics[width=0.95\columnwidth]{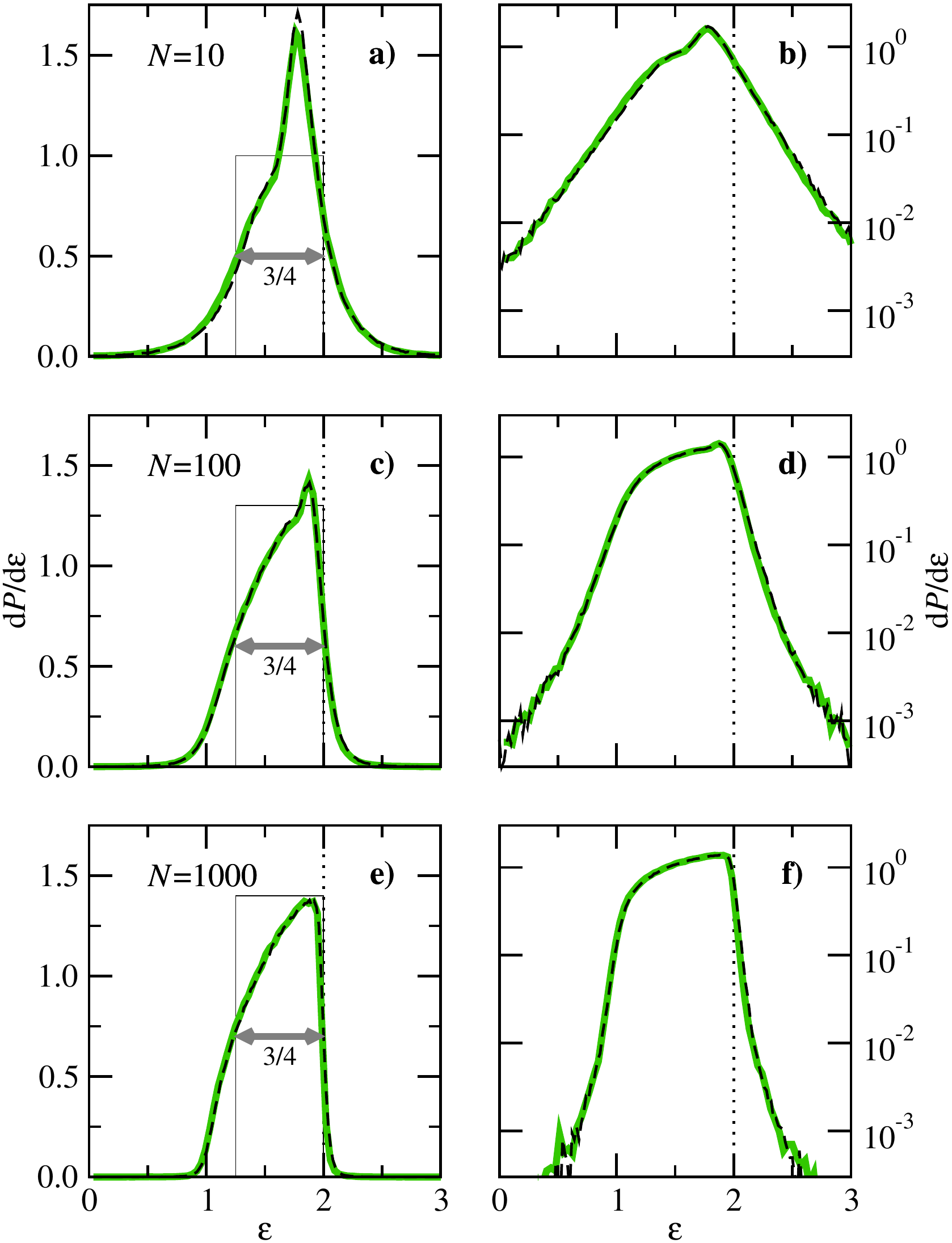}
\end{center}
\caption{Color online: Photo-electron spectra for propagated Coulomb complex (solid green/gray line) and BIS approximation (dashed black) according to \eq{eq:n2sum} for $N=10$ (upper row), $10^2$ (middle) and $10^3$ (lower) on a linear (left column) and a logarithmic (right) scale, respectively. 
The gray arrows indicate the theoretical value of $\Delta\eps=3/4$ according to \eq{eq:delte} for the full width at half maximum of the mean-field distribution shown by thin solid lines.
}
\label{fig:n2sum}
\end{figure}%
%%%%%%%%%%%%%% 
We consider for electron $i$ the binary collision with electron $j$, with initial positions and velocities (denoted with a prime) as in the Coulomb complex. 
Within the BIS approximation, the final energy of electron $i$ is then
\begin{equation}\label{eq:n2sum}
\eps_i=\eps_i' + \sum\limits_{j (\neq i)}^{N} \left( \eps_{ij} - \eps_{ij}' \right)
\end{equation}
with $\eps_i'=\eps^*$ by construction.
$\eps_{ij}'$ and $\eps_{ij}$ are the initial and final energies of electron $i$ due the interaction or collision with electron $j$, respectively, i.e., $\eps_{ij}'=v'_{i}{}^{2}/2+1/r'_{ij}$ with $v'_{i}=|\vec{v}'_{i}|$ and $r'_{ij}\equiv\left|\vec{r}'_{i}{-}\vec{r}'_{j}\right|$ for two electrons with initial positions $\vec{r}'_i$, $\vec{r}'_j$ and velocities $\vec{v}'_i$, $\vec{v}'_j$.
The final energy $\eps_{ij}$ can be calculated analytically by means of the conserved Runge-Lenz vector in the center-of-mass frame \cite{shgo+09}. Therefore we introduce the relative and center-of-mass coordinates $\vec{r}' = \vec{r}'_i {-} \vec{r}'_j$ and $\tilde{\vec{r}}' = (\vec{r}'_i {+} \vec{r}'_j)/2$, respectively. While the center-of-mass velocity is conserved ($\tilde{\vec{v}}=\tilde{\vec{v}}'$), we obtain an explicit expression for the final relative velocity $\vec{v}$ by means of the conserved angular momentum $\vec{l} \equiv \mu\, ( \vec{r}' \times \vec{v}' )$ and the Runge-Lenz vector $\vec{b} \equiv (\vec{v}' \times \vec{l} ) + \vec{r}'/r'$, with the reduced mass $\mu=1/2$. It reads $\vec{v} = -v \left(v \left( \vec{b} \times \vec{l} \right) - \vec{b} \right)/\left( 1 + v^2 l^2 \right)$, whereby the absolute value $v$ is known from energy conservation $v^{2} = 2/\mu r' + v'{}^{2}$.
Finally, we use $\eps_{ij}=\left(\tilde{\vec{v}}{+}\vec{v}/2\right)^{2}\!/2$.

The binary interaction does not explicitly include the influence of the background potential. 
However, it is easy to show that the definition \eqref{eq:n2sum} ensures conservation of the total energy $E$, which is, on the one hand, given by the \textsl{l.h.s.}\ of Eq.\,\eqref{eq:n2sum} $E=\sum_{i}\eps_{i}$.
On the other hand, it is $\sum_{i\ne j}^{N}(\eps_{ij}-\eps_{ij}')=-\sum_{i\ne j}^{N}1/r_{ij}$ from which follows that the \textsl{r.h.s.}\ of Eq.\,\eqref{eq:n2sum} is also equal to $E=\eps_i' - \sum_{i\ne j}^{N}1/r_{ij}$. Note that the BIS is computationally extremely cheap since it does not require numerical propagation. Yet, quite a few analytical binary collision outcomes need to be summed since all $N(N{-}1)$ pairwise interactions as well as multiple realizations of the isotropic velocity distribution need to be taken into account. Hereby, it is crucial that the direction of an electron's velocity for all its $N{-}1$ binary interactions within one realization of BIS is kept fixed, to ensure the correct particle-particle correlations of the binary energy exchanges. 

%%%%%%%%%%%%%%
\begin{figure}[t]
\begin{center}
\includegraphics[width=0.75\columnwidth]{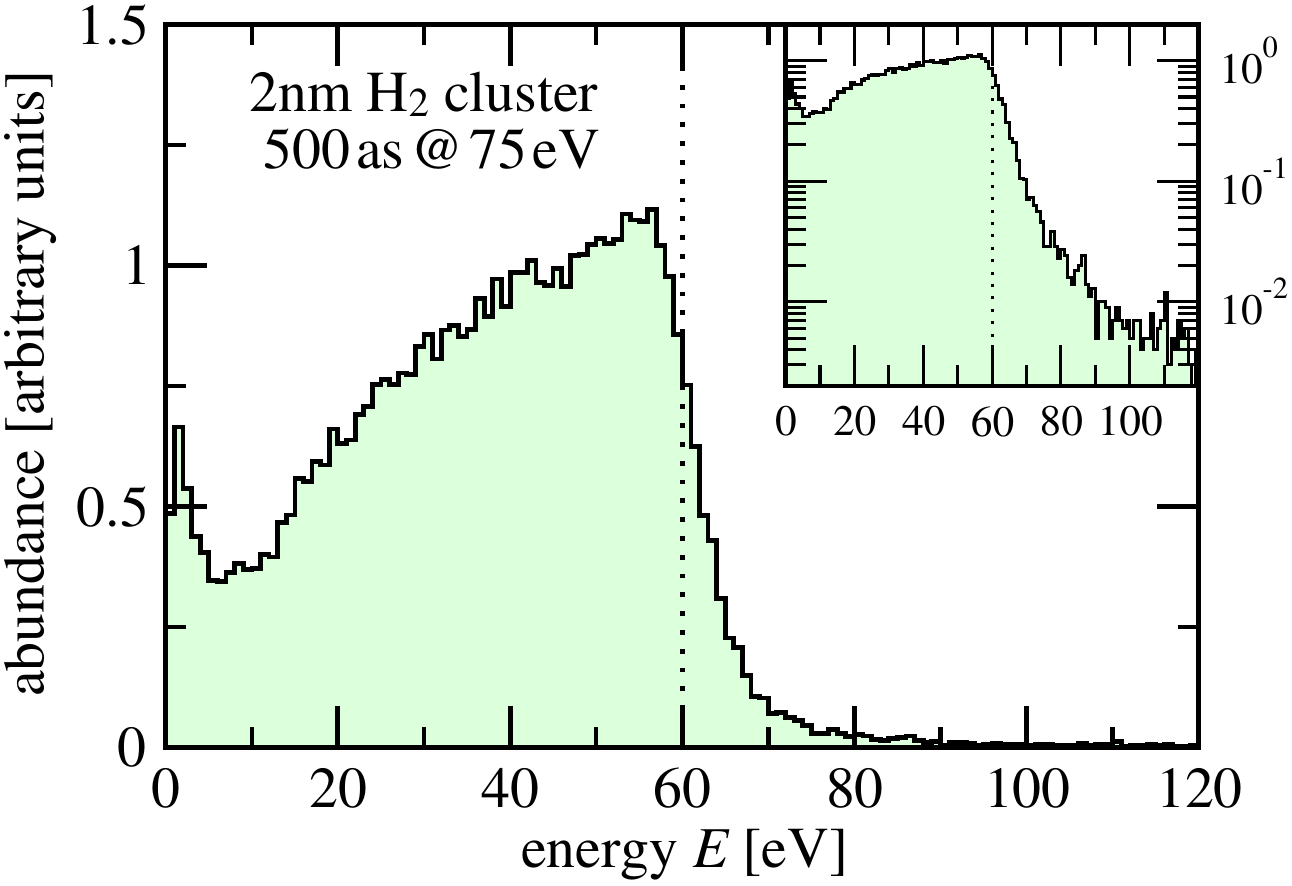}
\end{center}
\caption{Color online: Photo-electron spectrum from a 2\,nm H$_{2}$ cluster exposed to a pulse with 75\,eV photon energy, corresponding to $E^{\star}=60$\,eV (dotted vertical line). The pulse had an intensity of 2.5$\times$10$^{16}$W/cm$^{2}$ and a duration of 500\,as. The inset shows the same spectrum on logarithmic scale.
}
\label{fig:hydro}
\end{figure}%
%%%%%%%%%%%%%
The BIS spectra are shown in \fig{fig:n2sum}, overlayed with the fully propagated spectra. 
In all cases we see near perfect agreement. This agreement extends to specific features, such as a main peak arising from the residual discrete nature for the smaller systems with $N=10$ and $N=10^2$. Probably more surprisingly, also the long tails from violent collisions with very low probability are here reproduced in great detail (see the logarithmic spectra in the right column).
While these tails prevail for all system sizes shown, one sees a clear evolution towards mean-field dominated dynamics for larger systems with the central peak disappearing.
In fact the shoulder on the left wing for the two smaller systems indicates roughly the maximum due to mean-field dynamics.
This is corroborated by the width of the distribution $\Delta\eps$, which is in good agreement with the value in \eq{eq:delte} for all three cluster sizes if the shoulder of the spectrum is taken as the relevant maximum.

For the sake of clarity we have restricted ourselves to possible single (photo-)ionization of each constituent of the cluster or large molecule and to sudden photoionization to introduce the phenomenon of massively parallel ionization. That this is nevertheless a realistic scenario is demonstrated in \fig{fig:hydro} with the PES for a 2\,nm hydrogen cluster induced by an XUV pulse of 2.5$\times$10$^{16}$W/cm$^{2}$ peak intensity, a duration of 500\,as (full-width-at-half-maximum bandwidth of 4\,eV) and a photon energy of 75\,eV. This corresponds to an excess energy of $E^{\star}=60$\,eV (dotted vertical line in \fig{fig:hydro}).
Under these conditions, within reach by modern FEL sources, about 10\,\% of the about 500 molecules in the cluster are ionized.
The spectrum clearly shows the typical features for massively parallel ionization: a square-root shaped rise for energies $E<E^{\star}$ and a high-energy tail for $E>E^{\star}$, see inset of \fig{fig:hydro}. 
At very low energies one observes a structure due to electron-impact ionization which would increase for larger clusters.

To summarize, we have introduced the phenomenon of massively parallel ionization which is a so far unexplored variant of multi-photon ionization and should routinely occur when illuminating larger targets with intense and short XUV to X-ray pulses. Characteristic for massively parallel ionization is a photo-electron spectrum which combines, almost independently, 
features from mean-field dynamics with those of violent binary electron collisions. Thereby, we could demonstrate that one can deduce from the width of the spectrum the ratio of charging (number of ionized electrons) to the size (radius) of the systems which is of great diagnostic value in experiments. Moreover,
we have devised a quasi-analytical yet highly accurate method to calculate the photo-electron spectrum from a sum of synthetic binary Coulomb collisions for small to large systems.
Clearly, depending on photon energy and target, Auger-decay processes and multiple photo-ionization of one atom or small molecule within the cluster can occur and will modify the results presented here. How the characteristic features of massively parallel ionization will be changed by such events will be investigated in future work.


\begin{thebibliography}{10}

\bibitem{acas+07}
W. Ackermann {\it et~al.}, Nat. Photon. {\bf 1}, 336 (2007).

\bibitem{emak+10}
P. Emma {\it et~al.}, Nat. Photon. {\bf 4}, 641 (2010).

\bibitem{sasi+06}
U. Saalmann, C. Siedschlag, and J.~M. Rost, J. Phys. B {\bf 39}, R\,39
 (2006).

\bibitem{newo+00}
% R. Neutze, R. Wouts, D. van~der Spoel, E. Weckert, and J. Hajdu,
R. Neutze {\it et~al.},
 Nature {\bf 406}, 752 (2000).
 
\bibitem{both+10}
C. Bostedt {\it et~al.}, New J. Phys. {\bf 12}, 083004 (2010).

\bibitem{arfe10}
M. Arbeiter and T. Fennel, Phys. Rev. A {\bf 82}, 013201 (2010).

\bibitem{gnsa+11}
C. Gnodtke, U. Saalmann, and J.-M. Rost, New J. Phys. {\bf 13}, 013028
 (2011).

\bibitem{arfe11}
M. Arbeiter and T. Fennel, New J. Phys. {\bf 13}, 053022 (2011).

\bibitem{sage+08}
U. Saalmann, I. Georgescu, and J.~M. Rost, New J. Phys. {\bf 10}, 025014
 (2008).

\bibitem{both+08}
C. Bostedt {\it et~al.}, Phys. Rev. Lett. {\bf 100}, 133401 (2008).

\bibitem{mo09}
K. Moribayashi, Phys. Rev. A {\bf 80}, 025403 (2009).

\bibitem{saro02}
U. Saalmann and J.~M. Rost, Phys. Rev. Lett. {\bf 89}, 143401 (2002).

\bibitem{jufa+04}
Z. Jurek, G. Faigel, and M. Tegze, Eur. Phys. J. D {\bf 29}, 217 (2004).

\bibitem{shgo+09}
N.~I. Shvetsov-Shilovski, S.~P. Goreslavski, S.~V. Popru\-zhenko, and W. Becker,
 Laser Phys. {\bf 19}, 1550 (2009).

\end{thebibliography}
\end{document}